 \documentclass[final,5p,times,twocolumn]{elsarticle}

\usepackage{amssymb, bm}
\usepackage{lipsum}
\usepackage{xcolor}
\usepackage{hyperref}

\newcommand{\appropto}{\mathrel{\vcenter{
  \offinterlineskip\halign{\hfil$##$\cr
    \propto\cr\noalign{\kern2pt}\sim\cr\noalign{\kern-2pt}}}}}

\journal{Journal of Magnetism and Magnetic Material, and accepted on  Oct 4, 2023: https://doi.org/10.1016/j.jmmm.2023.171359 }

\begin{document}

\begin{frontmatter}



\title{Unidirectional spin wave emission  by travelling pair of magnetic field profiles}


\author[inst1]{Gauthier Philippe}
\author[inst1]{Mathieu Moalic}
\author[inst1]{Jarosław W. Kłos}
\ead{klos@amu.edu.pl}

\affiliation[inst1]{organization={ISQI, Faculty of Physics, Adam Mickiewicz University},%Department and Organization
            addressline={Uniwersytetu Poznańskiego 2}, 
            postcode={61-614}, 
            city={Poznań},
            country={Poland}}

\begin{abstract}
We demonstrate that the spin wave Cherenkov effect can be used to design the unidirectional spin wave emitter with tunable frequency and switchable direction of emission. In our numerical studies, we propose to use a pair of traveling profiles of the magnetic field which generate the spin waves, for sufficiently large velocity of their motion. In the considered system, the spin waves of shorter (longer) wavelengths are induced at the front (back) of the moving profiles and interfere constructively or destructively, depending on the velocity of the profiles. Moreover, we showed that the spin waves can be confined between the pair of traveling profiles of the magnetic field. 
This work opens the perspectives for the experimental studies in hybrid magnonic-superconducting systems where the magnetic vortices in a superconductor can be used as moving sources of the magnetic field driving the spin waves in the ferromagnetic subsystem.
\end{abstract}

%\includegraphics{TOC_figure.pdf}

%\item Cherenkov emission of spin waves was numerically %investigated.
%\item The spin waves are emitted unidirectionally for selected %velocities of the Cherenkov sources.
%\item The moving Abrikosov vortices can be considered as sources %of spin waves.
%\clearpage




\end{frontmatter}

%\tableofcontents



\section{Introduction}

\label{introduction}

The Cherenkov and Doppler effects are the fundamental wave phenomena resulting
from a uniform motion of the sources~\cite{Ginzburg_1996}. The Doppler effect
is related to the change of the frequency of (monochromatic) wave source
$\omega\rightarrow\omega'$~\cite{Perrine_1944} due to its motion with constant velocity
$\boldsymbol{v}$: $\omega'=\gamma\left(\omega+\boldsymbol{v}\cdot \boldsymbol{k}(\omega)\right)$~\cite{Berger1976}, where  $\gamma=1$ or $1/\sqrt{1+(v/v_\varphi)^2}$,
depending if the transformation between reference frames is described by
Galilean or Lorentz transformation. The Cherenkov effect~\cite{Smith1993} is
observed as a generation of the waves by the source moving with the velocity
$v$ equal to or larger than the phase velocity  $v_\varphi$ of the
medium:  $v\ge v_\varphi$. It is worth noting that this effect exists even if the
`source' is `static' in moving reference frame: $\omega=0$. In this case, the
equation $\omega'(\boldsymbol{k}) = \boldsymbol{v}\cdot\boldsymbol{k}$ determines the frequency(ies) $\omega'$ and the
corresponding wave vector(s) $\boldsymbol{k}$ of excited waves, which is(are)
related by the condition: $v_\varphi=\omega'/k=v$.

The Cherenkov effect was observed for the first time in 1934 when the
$\gamma$-radiation emitted by pure liquids under the action of fast
electrons ($\beta$ --- particles of radioactive elements) was
detected~\cite{Cerenkov_1937}. The condition $v > v_\varphi$ can be fulfilled
because the velocity of the emitted electrons ($v\simeq c$) exceeds the phase
velocity of light in material medium $v_\varphi=c/n$ of the refractive index
$n>1$. The first theoretical explanation of the Cherenkov effect was
presented by I. Tamm and I. Frank~\cite{Frank1991} in the late thirties. 
 Nowadays, the Cherenkov effect is the subject of intensive studies not only in
the field of high-energy physics but also in condensed matter, and in
particular in photonics~\cite{Luo_2003,Xi_2009} and derived field:
polaritonics~\cite{Liu2012,genevet_2015,Rivera_2020,Zhang_2020}. It is worth
mentioning that electromagnetic waves are not the only platform on which the
Cherenkov effect can be studied and used in nanodevices.
%\QUERY[4]
Magnonics~\cite{Chumak_2022} offers equally interesting possibilities. The
phase velocities of spin waves are on the order of  few km/s, making the
Cherenkov effect relatively easy to observe.

Ten years ago, M. Yan~\cite{Yan2011,Yan2013} demonstrated numerically that
Cherenkov effects for spin waves can be excited by the moving pulse of the
magnetic field. The authors also found the formation of the Mach cones for 2D
and 3D ferromagnetic systems. The experimental realization of this idea is
challenging because it requires the generation of the fast-moving profile
(barrier) of the magnetic field. Such motion can be approximated, in a
time-lapse manner, by sequential application of the voltage to the long
sequence of the electrodes deposited on the magnetic layer in which we can
induce the magnetocrystalline anisotropy (and related effective
field)~\cite{Rana_2019}. Another approach, which is now intensively studied, is
based on the motion of fluxons in the superconducting layer. The fluxons
produce a stray field and can be pushed through a superconductor with large
velocities~\cite{dobrovolskiy_2020,dobrovolskiy_2019}. It was already
experimentally demonstrated that moving fluxons can induce a Cherenkov
radiation of spin waves in the ferromagnetic layer underneath the
superconductor~\cite{dobrovolskiy2023}. 

The uniform motion of the medium leads also to Doppler or Cherenkov effect.
This effect is well known in acoustics and has practical application in
ultrasonography~\cite{Evans2000}. The corresponding effect is observed in
magnonics if the spin wave is accompanied by the spin current flowing through
the system~\cite{Rossier2004} --- i.e.,~the precessional dynamic of
magnetization takes place on the top of the uniform motion of magnetic moments.
In such systems, one can observe Doppler~\cite{Vlaminck2008} of Cherenkov
effect~\cite{chen2016micromagnetic,Kruijf2017} for spin waves.

In our work, we do not consider the flow of spin current, but we focus on the
spin wave generation by the motion of linear barriers of the magnetic field.
Such a barrier, moving with a constant velocity, generates spin waves both in
the forward and backward direction, with respect to the direction of the
barrier's motion. The forward and backward propagating spin waves differ in the
wavelength~\cite{Yan2011}, which makes the considered spin wave emitter
non-reciprocal with a change in the direction of its motion. We propose to use
pair of such barriers, which move in parallel, to construct the unidirectional
spin wave emitter. Research on unidirectional spin wave emitters is being
carried out by many groups~\cite{Krivoruchko2018,Gallardo2021}. The proposed
system makes it possible to control the direction of spin wave propagation
(forward or backward) by tuning the velocity of the profile. Moreover, we can
block the emission of spin waves by confining them between moving barriers.

The article is organized as follows. After the introduction, we describe the
system under consideration and present the principle of operation of the
unidirectional emitter. Then, we briefly introduce the applied model and the
computational technique. In the next section, we present the results for a
single barrier~\cite{Yan2011}, which is a reference system in our studies.
After that, we discuss the outcomes for a pair of barriers illustrating three
scenarios: forward emission, backward emission, and spin wave confinement. The
work concludes with a summary.

\section{Structure and model}
\label{sec:structure}

\begin{figure}[h]
    \centering 
    \includegraphics[width=1.0\columnwidth]{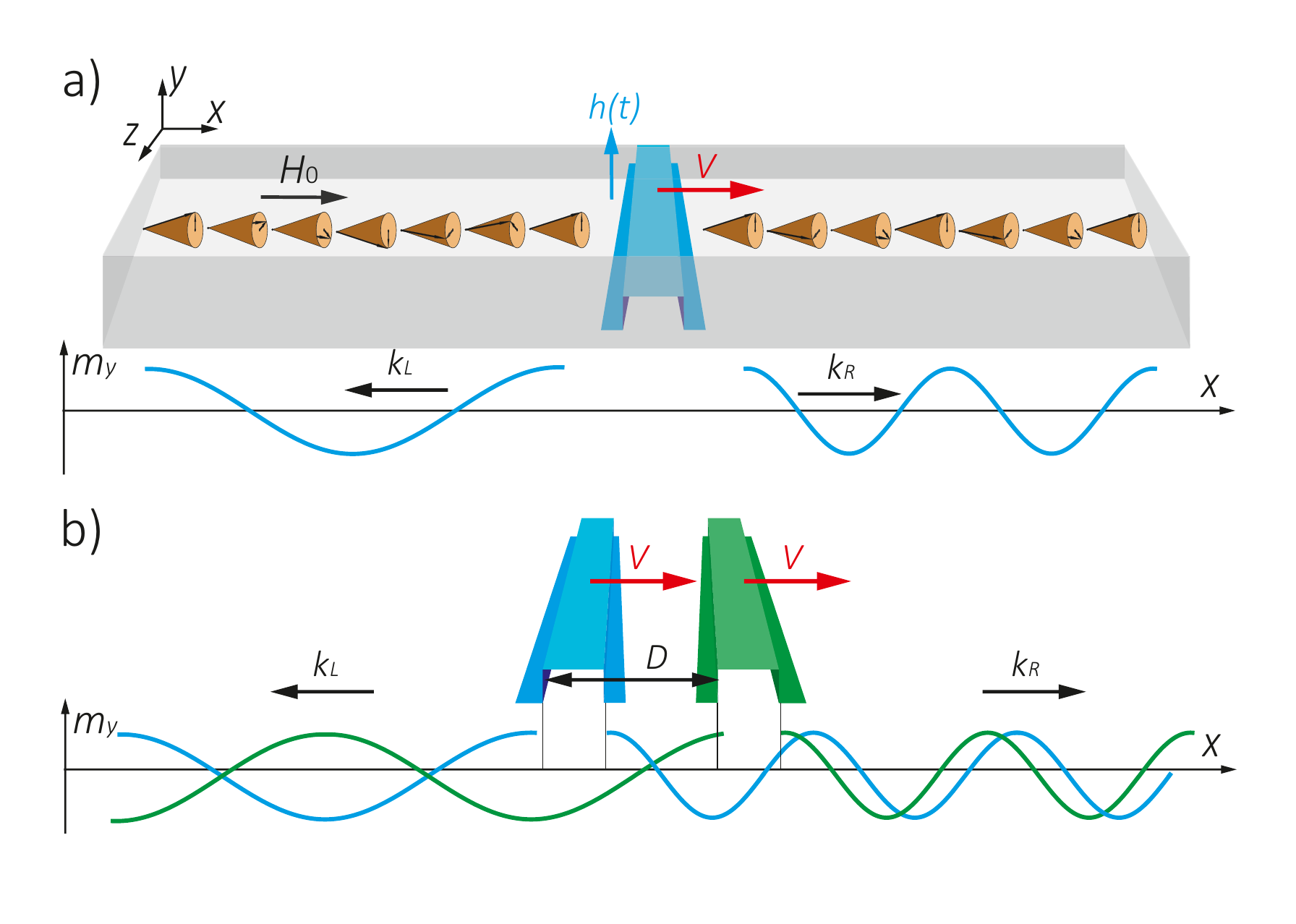}	
        \caption{(a)  The ilustration of the spin wave Cherenkov effect. The spin waves of different wave vectors $\boldsymbol{k}_L$ and $\boldsymbol{k}_R$ are excited by the linear profile of the magnetic field of rectangular cross-section $h(x=v t)=h_0\,\theta(x+d/2)\,\theta(d/2-x)$, moving uniformly with the velocity $v_0$, where $d$ and $h_0$ denote the width and height of the profile, and $\theta(x)$ is a unit step. The magnetic layer (grey box) is magnetized in-plane, and the field $H_0$ is applied along the profiles's motion, which is parallel to the direction of spin waves' propagation: $\boldsymbol{k}_L$ (and $\boldsymbol{k}_R$). (b) The principle of working for unidirectional spin wave emitter. The spin waves produced by two parallelly moving barriers  (separated by fixed distance $D$)  can interfere destructively or constructively on the opposite sides of the moving barriers. } 
         \label{fig:system}
\end{figure}
\begin{figure} 
\centering
    \includegraphics[width=0.5\textwidth]{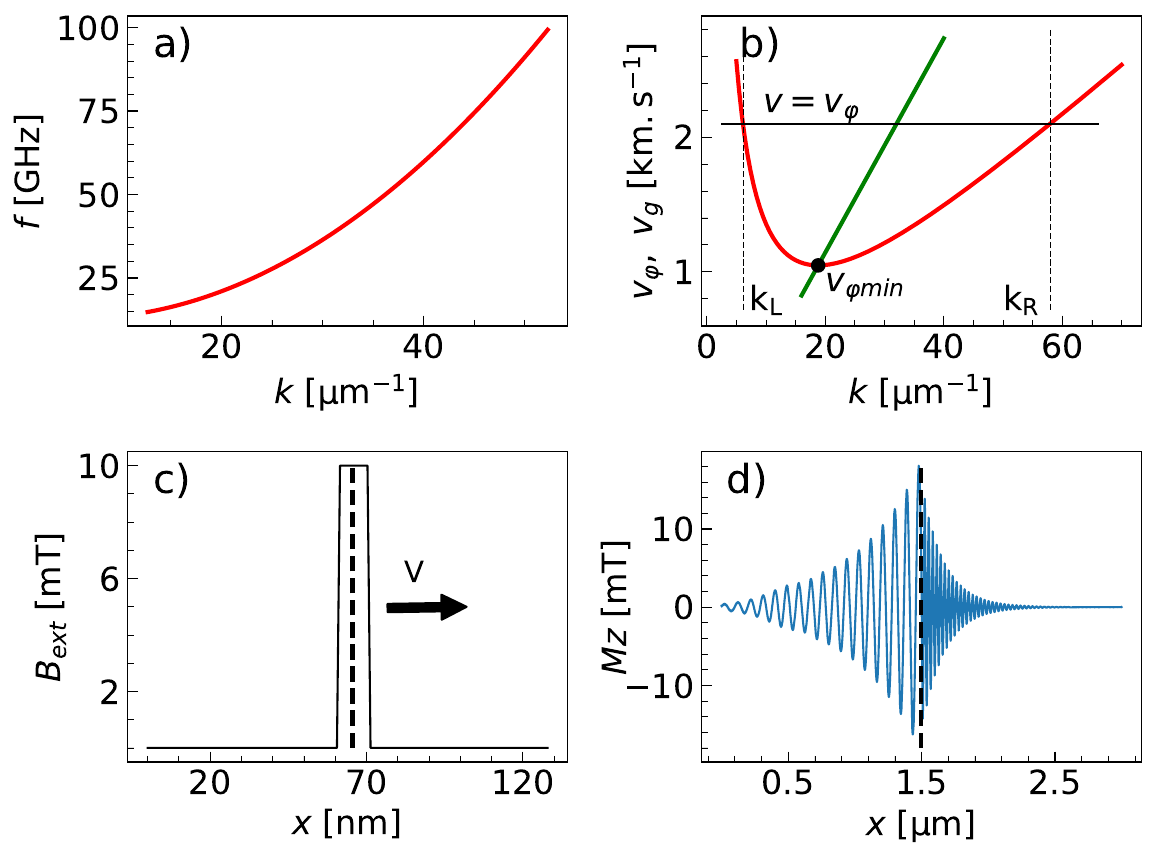}
    \label{fig:spectrum}
    \caption{(a) The dispersion relation for considered magnetic film in backward volume geometry: frequency versus wave vector $f(k)$, and (b) the related dependences of phase velocity $v_\varphi$ (red line) and group velocity $v_g$ (green line) on the wave vector. The horizontal black line denotes (c) the velocity $v$ of the square profile of the magnetic field. The velocity $v_0$ determines the wave numbers $k_L$ and $k_R$ of (d) the spin waves emitted backward and forward, respectively, generated by the moving profile of the magnetic field.
     %   Result are the wavevector of the waves generated by the spin Cherenkov effect for an excitation velocity, V$\varphi$ and Vg are the phase velocity and group velocity get from the dispertion relation,  \\c) Moving excitation oriented to the z axis, \\d) Magnetization of the system for V=1300 m/s
     } 
\end{figure}

It is known~\cite{Yan2011,Yan2013} that fastly moving profile of a magnetic
field can generate spin waves, which differ in the wavelength, depending on the
direction of propagation (see  \ref{fig:system}(a)). This effect is known as a
spin wave Cherenkov effect. Interestingly, the wavelength (wavenumber) of the
forward and backward propagating spin waves changes with different rates as the
velocity $v$ of the barrier increases (see  \ref{fig:spectrum}(b)).
This allows designing \textit{the unidirectional spin wave emitter} where the
spin waves produced by the pair of moving profiles of the magnetic field can
interfere constructively or destructively on the opposite sides of the system
--- see  \ref{fig:system}(b). The conditions for the observation of the
constructive (and destructive) interference in the front (and in the back) are
not accidental and can be tuned by the adjustment of the selection of the
velocity $v$.

We considered a ferromagnetic stripe with a thickness of 10~nm and a width of
100~nm as a conduit for spin waves, which has been magnetized alongside the
external field $H_0\mu_0=1 ~{\rm T}$. It means the backward volume configuration for spin
waves where their wave vector is parallel to the external field. We assumed
that the ferromagnetic material (parmalloy) is characterized by the saturation
magnetization $M_{\rm S}= 796\times10^{3} ~ {\rm A/m} $,  exchange stiffness $A_{\rm ex}= 1.3\times10^{-11}~ {\rm J/m}$, and the low damping
$\alpha=0.02$~\cite{Yan2013}. On both ends of the stripe, we implemented
absorbing boundary conditions by gradually increasing the value of $\alpha$. 

We used a modified version of Mumax3~\cite{mumax_2014}, the GPU-accelerated
micromagnetic software, which solves the Landau--Lifshitz--Gilbert equation to
simulate the magnetization dynamics. To calculate the spin wave dispersion, we
applied the harmonic (in time) and  \textit{sinc}-shaped (in space) pulse of
magnetic field on one side of the magnetic stripe. We assumed the cut-off wave
number $k_{\rm cut}=1\times 10^{8}$~1/m and sweeped the frequency $f$ starting from 14
to 30~GHz by the steps of 0.5~GHz and from 30 to 100~GHz by steps of 1~GHz.
After the time $50/f$, we recorded the spin wave on the opposite side of
the wire for each step of the simulations. The recorded spin wave profile was
post-processed, using Fourier transform, to determine the leading wave vector
corresponding to a given frequency. To observe the spin wave Cherenkov effect,
we generated the moving profile of a magnetic field of rectangular shape
($d=10$ nm in width and $h_0$ = 10 mT in height) --- see
\ref{fig:spectrum}(c). We registered the spin waves excited by the moving
profile of the magnetic field, for successive values of its velocity. The
simulations were performed for different velocities of magnetic profile
$v$ ranging from  $\mathrm{500~m/s}$ to  $\mathrm{2500~ m/s}$.  Such values are
observed experimentally for moving sources of magnetic field in the form of
superconducting vortices~\cite{Dobrovolskiy2023(b)} in hybrid
superconducting-magnonic systems~\cite{Dobrovolskiy2019(b)}.

\section{Cherenkov radiation of spin waves}

The Cherenkov effect for electromagnetic waves is usually associated with the radiation which occurs when a charged particle moves through a material with a higher velocity than the material's phase velocity for light. When the charged particle moves with a velocity smaller than the phase velocity of light, there is a deformation of the electric polarization in the material around the charged particle. In the reverse situation, when the velocity of the charged particle is larger than the phase velocity in the medium, the deformation of the electric field does not have the time to recover its initial state, so the deformation is extended on the particle trajectory, and creates an electromagnetic wave.

A similar effect is observed for spin waves when the magnetization is locally modified by the moving magnetic excitation (narrow profile of magnetic field).  If the velocity of the excitation $v$ exceeds (the minimum value) of the phase velocity in the magnetic medium $v>v_{\varphi_{\rm min}}$, the magnetization does not have the time to recover its initial state in the time of the flight of excitation and a spin wave is generated --- see  \ref{fig:system}(a) and  \ref{fig:spectrum}(b).

In our study, we are going to demonstrate that the pair of the profiles of the
magnetic field moving parallelly at properly selected velocities can work as a
unidirectional spin wave emitter --- see  \ref{fig:system}(b). To test our
numerical model and illustrate the principles of the spin wave Cherenkov
effect, we reproduced the result of M.~Yan~\cite{Yan2011,Yan2013}, where the
motion of a single profile of magnetic field was considered.

 \ref{fig:spectrum}(a) presents the numerically determined dispersion relation $f(k)$ (frequency versus wave number) for the considered stripe (see  \ref{sec:structure}). From the relation $f(k)$, we calculated the dependence of the spin wave phase (and group) velocity $v_\varphi$ ($v_g$) on the wave number $v_\varphi=2\pi f/k$ ($v_g=2\pi\,df/dk$) --- see  \ref{fig:spectrum}(b). It is interesting to notice that the system has a threshold value of the phase velocity for spin waves, corresponding to the minimum $v_{\varphi_{\rm min}}$ of $v_\varphi(k)$. According to the condition $v=v_\varphi$ describing the spin wave Cherenkov emission, the spin waves cannot be generated when $v<v_{\varphi_{\rm min}}$ and for $v>v_{\varphi_{\rm min}}$, the spin waves of two different wave numbers (and corresponding frequencies) are emitted. The minimum of $v_{\varphi_{\rm min}}$ corresponds to the condition: $dv_\varphi(k)/dk=0\;\Rightarrow v_\varphi=\frac{1}{2\pi}dv_\varphi/df=v_g$. Therefore, a wave with a smaller (larger) wave number will propagate with the slower (faster) that the field's profile $v_g<v$ ($v_g>v$) and remain behind (overtake) the moving field's profile.

\begin{figure}[h]
\centering
    \includegraphics[width=0.48\textwidth]{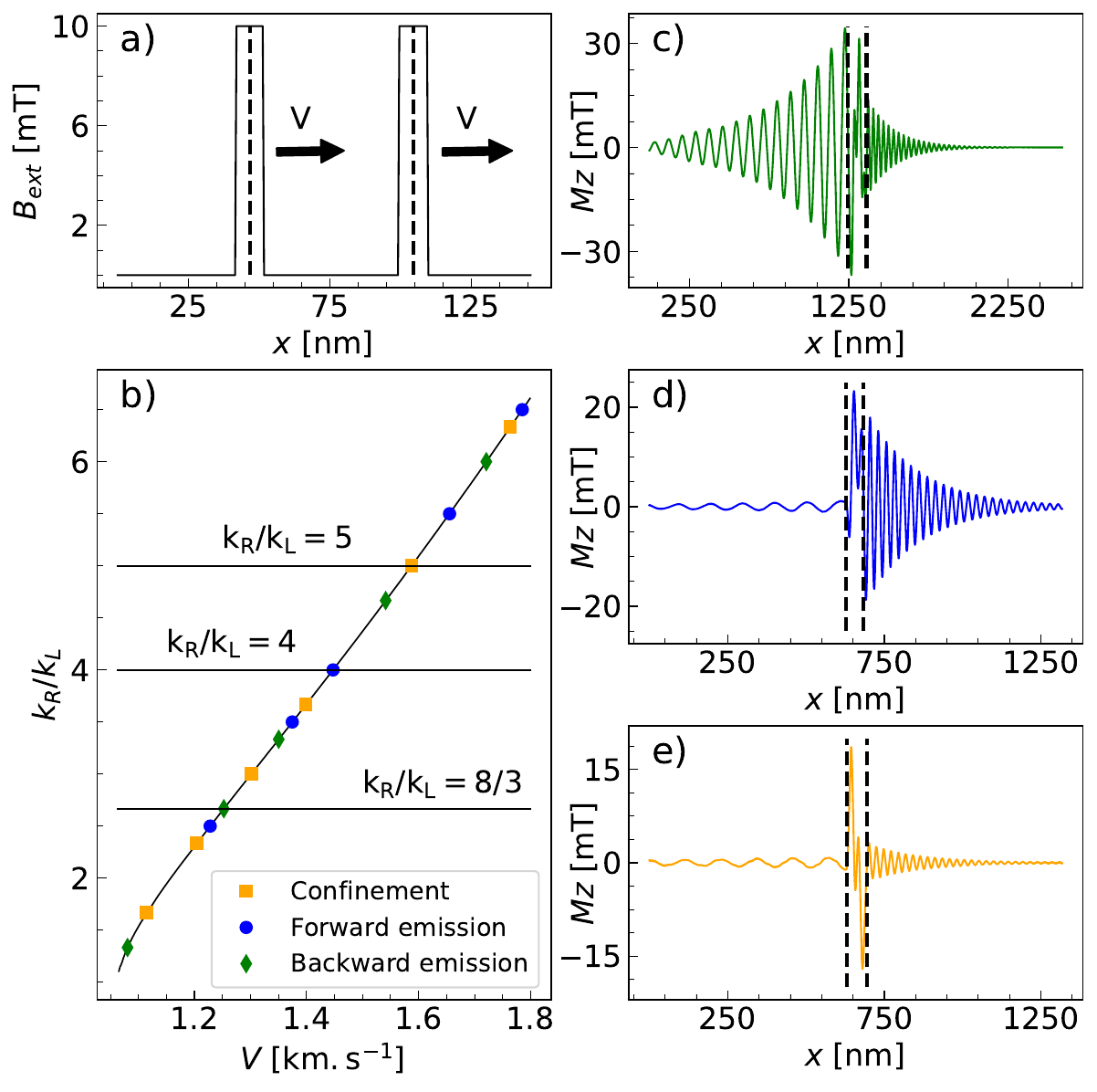}
    \label{fig:emitter}
    \caption{(a) Two profiles of the magnetic field which move with the same velocity $v_0$, keeping fixed distance $D$. (b) The ratio of the wave numbers $k_R/k_L$ for different values of the velocity $v_0$; square, circular, and rhombic dots mark the  $k_R/k_L$ ratios, for which we observe: (c) unidirectional backward spin wave emission, (d) unidirectional forward emission, (e) partial confinement between the field profiles. }
\end{figure}

\section{Tunable, unidirectional spin wave emitter}

  Let us discuss now the working principles of unidirectional spin wave emitter presented in  \ref{fig:system}(b) where two square profiles of the magnetic field move with the same speed $v$, keeping a constant gap $D$ between them (\ref{fig:system}(b) and  \ref{fig:emitter}(a)).

To observe destructive (constructive) interference of two harmonic sources generating the waves of the wavevector $k$ and displaced by the distance $D$, the wave number should fulfill the following condition: $k D = 2\pi n$ ($k D = \pi(2n+1)$), where $n$ is an integer number. In the considered system, we can tune the value of the wave vector of the generated spin wave $k(v_\varphi=v)$ by changing the velocity $v$ of the moving field's profile. It is worth noting that this tuning $k(v)$ takes place with a different rate for forward propagating spin waves (of larger wavenumber $k_R$) and backward propagating spin waves (of smaller wave  $k_L$). It is known that the dipolar-exchange dispersion relation $f(k)$ is linear: $f\propto k$ (quadratic $f\propto k^2$) for small (large) wave numbers. This corresponds to the relation: $v_\varphi\propto 1/k$ ($v_\varphi\propto k$) for small (large) wave numbers. As a result, the ratio of $k_L(v)/k_R(v)$ will vary approximately linearly as the velocity $v=v_\varphi$ increases --- see  \ref{fig:emitter}(b). We can consider three particular scenarios.
\begin{itemize}
    \item \emph{Backward spin wave emitter} -- constructive interference in the back and destructive interference in the front of moving barriers --  \ref{fig:emitter}(c):
    \begin{equation}
        {\frac{k_L(v)}{k_R(v)} = \frac{2n_L}{2n_R+1}\appropto v.}\label{eq:SW_beck}
\end{equation}
    \item \emph{Forward spin wave emitter} -- constructive interference in the back and destructive interference in the front of moving barriers --  \ref{fig:emitter}(d):
    \begin{equation}
        {\frac{k_L(v)}{k_R(v)} = \frac{2n_L+1}{2n_R}\appropto v.}\label{eq:SW_front}
\end{equation}
        \item \emph{Spin wave confinement} -- constructive interference in the back and destructive interference in the front of moving barriers --  \ref{fig:emitter}(e):
    \begin{equation}
        {\frac{k_L(v)}{k_R(v)} = \frac{2n_L+1}{2n_R+1}\appropto v.}\label{SW:conf}
\end{equation}
\end{itemize}
The symbols $n_L$ and $n_R$ are two independent integer numbers.

For larger velocities $v$ (i.e.~for $v>1.2$ km/s), the ratio $k_L(v)/k_R(v)$ is proportional to $v$. As the velocity of the profiles $v$ increases, it can be tuned multiple times to every three mentioned scenarios. In is worth noting that the ratio $k_L/k_R$ must be greater that one: $k_L/k_R>1$, which corresponds to the condition for Cherenkov emission $v>v_{\varphi_{min}}$.

Because of the damping, the constructive and destructive interferences cannot
be perfect. However, the effects of unidirectional emissions are quite
distinctive  --- see  \ref{fig:emitter}(a,b). The interfering waves decay
exponentially $e^{-x/l}$ with the rate $l=v_{\rm gr}\tau\appropto k^{-1}$ for the exchange dominated regime (and dipolar dominated regime), where the relaxation time is: $\tau=-\Im(\omega)=\gamma \mu_0 \;\omega^{-1}\; ({\partial\omega}/{\partial H_{0,{\rm eff}}})^{-1}\propto k^{-2}$ ($\propto k^{-1}$) and the group
velocity: $v_{\rm gr}\propto k$ ($\propto$ {\em const}$(k)$)~\cite{Stancil2009} --- $H_{0,{\rm eff}}$ is the static effective
field, here: $H_{0,{\rm eff}}=H_0$. Then the amplitude of the interfering waves will
change as $e^{\pm x k}(1\pm C e^{\pm D k})$, where $C$ is constant, the -(+) signs in the
exponents refer to the right(left) propagating waves, and the -(+) sign between
the terms in the brackets denotes the destructively(constructively) interfering
waves. This means that destructively interfering waves are not canceled out
immediately. This effect and the overall decay of interfering waves depends on
the $k$-number. However, the lack of perfect unidirectional emission
and confinement cannot be attributed solely to the $k$-dependent
damping, but is possibly related to the occurrence of nonlinear effects in this
externally pumped system.

\section{Summary}

 Our simulations demonstrate that it is possible to use the spin wave Cherenkov emission to design the unidirectional (backward or forward) spin wave emitter of tunable frequency.

We showed that it is feasible to confine and continuously pump the bi-harmonic superposition of spin waves (i.e.~the spin waves of two different frequencies).

The discussed effects can be potentially implemented in hybrid magic-superconducting systems where the Abrikosov lattices vortices can be used as moving sources of the magnetic field that drives the spin waves in the ferromagnetic subsystem.

\section*{Data availibility}
All of the simulations scripts, simulation outputs and post-processing code is open source and has been made available at \href{https://doi.org/10.5281/zenodo.8328602
}{https://doi.org/10.5281/zenodo.8328602}

\section*{Acknowledgements}
G. P. anf J. W. K. would like to acknowledge the erasmus mundus MaMaSELF programm  and the support from the National Science Center – Poland grant No. 2021/43/I/ST3/00550.

\appendix

%\section{Appendix title 1}

%\section{Appendix title 2}

%\bibliographystyle{elsarticle-harv} 

%\bibliographystyle{elsarticle-num}
%\bibliography{bibliography}

\end{document}